\documentclass[10pt,conference]{IEEEtran}

\usepackage{import}
\usepackage[numbers,compress]{natbib}

\usepackage{mathptmx}   
\usepackage{emptypage}  
\usepackage{blindtext}  
\usepackage{ragged2e}   
\usepackage{pdflscape}  
\usepackage{multicol}   
\usepackage{gensymb}    
\usepackage{bold-extra} 
\usepackage{amsmath}    
\usepackage{amssymb}    
\usepackage{hyperref}   
\usepackage{cleveref}   

\makeatletter
\let\MYcaption\@makecaption
\makeatother

\usepackage[font=footnotesize]{subcaption}

\makeatletter
\let\@makecaption\MYcaption
\makeatother

\usepackage{enumitem}
\usepackage[multiple]{footmisc} 

\newcommand{\footnoteurl}[2]{\footnote{{#1} last accessed #2.}} 
\usepackage{graphicx}
\graphicspath{{images/}{../images/}}

\usepackage{listings}
\usepackage{textcomp}
\usepackage{color}
\definecolor{maroon}{rgb}{0.5,0,0}
\definecolor{darkgreen}{rgb}{0,0.5,0}

\newcommand{\beginlstdelim}[3]
{%
  \def\endlstdelim{#2\egroup}%
  \ttfamily#1\bgroup\color{#3}\aftergroup\endlstdelim%
}

\lstdefinestyle{xml_style}
{
  morestring=[b]",
  moredelim=*[s][\bfseries\color{maroon}]{<}{>},
  moredelim=*[s][\bfseries\color{maroon}]{</}{>},
  morecomment=[s]{<?}{?>},
  morecomment=[s]{<!--}{-->},
}

\lstdefinestyle{json_style}
{
  morestring=[b]",
  comment=[l]{//},
  keywordstyle=\bfseries\color{maroon}
}

\lstdefinelanguage{JSON}
{
  style=json_style
}

\lstdefinelanguage{XML}
{
  style=xml_style
}

\usepackage{booktabs}
\usepackage{multirow}
\usepackage{makecell}

\usepackage{xcolor}

\IEEEoverridecommandlockouts 

\usepackage{balance}

\usepackage{framed}

\def \dima{Usage Description}
\def \dimb{Design Rationale}
\def \dimc{Domain Concepts}
\def \dimd{Support Artefacts}
\def \dime{Documentation Presentation}

\begin{document}

\title{
What should I document? A preliminary systematic mapping study into API documentation knowledge
}

\author{
  \IEEEauthorblockN{Alex Cummaudo}
  \IEEEauthorblockA{
    \textit{Applied Artificial Intelligence Institute} \\
    \textit{Deakin University}\\
    Geelong, Australia \\
    ca@deakin.edu.au
  }
  \and
  \IEEEauthorblockN{Rajesh Vasa}
  \IEEEauthorblockA{
    \textit{Applied Artificial Intelligence Institute} \\
    \textit{Deakin University}\\
    Geelong, Australia \\
    rajesh.vasa@deakin.edu.au
  }
  \and
  \IEEEauthorblockN{John Grundy}
  \IEEEauthorblockA{
    \textit{Faculty of Information Technology} \\
    \textit{Monash University}\\
    Clayton, Australia \\
    john.grundy@monash.edu
  }
}

\maketitle

\IEEEpubidadjcol 
\begin{abstract}

\textit{Background}: 
Good API documentation facilities the development process, improving productivity and quality.
While the topic of API documentation quality has been of interest for the last two decades, there have been few studies to map the specific constructs needed to create a good document. In effect, we still need a structured taxonomy against which to capture knowledge. 
\textit{Aims:}
This study reports emerging results of a systematic mapping study. We capture key conclusions from previous studies that assess API documentation quality, and synthesise the results into a single framework.
\textit{Method:}
By conducting a systematic review of 21 key works, we have developed a five dimensional taxonomy based on 34 categorised weighted recommendations.
\textit{Results:}
All studies utilise field study techniques to arrive at their recommendations, with seven studies employing some form of interview and questionnaire, and four conducting documentation analysis.
The taxonomy we synthesise reinforces that usage description details (code snippets, tutorials, and reference documents) are generally highly weighted as helpful in API documentation, in addition to design rationale and presentation.
\textit{Conclusions:}
We propose extensions to this study aligned to developer's utility for each of the taxonomy's categories.

\end{abstract}

\section{Introduction}

Improving the quality of API documentation is highly valuable to the software development process; good documentation facilitates productivity and thus quality is better engineered into the system \citep{mcleod2011factors}. Where an application developer integrates new pieces of functionality (via APIs) into a system, their productivity is affected either by inadequate skills (\textit{``I've never used an API like this, so must learn from scratch''}) or, where their skills are adequate, an imbalanced cognitive load that causes excessive context switching (\textit{``I have the skills for this, but am confused or misunderstand''}). In the latter case, what causes this confusion and how to mitigate it via improved API documentation is an area that has been explored; prior studies have provided recommendations based on both qualitative and quantitative analysis of developer's opinions. These recommendations and guidelines propose ways by which developers, managers and solution architects can construct systems better.

However, to date there has been little attempt to systematically capture this knowledge about API documentation from various studies into a readily accessible, consolidated format, that assists API designers to prepare better documentation. While previous works have covered certain aspects of API usage, many have lacked a systematic review of literature and do not offer a taxonomy to consolidate these guidelines together. For example, some studies have considered the technical implementation improving API usability or tools to generate (or validate) API documentation from its source code (e.g., \citep{Nybom:2018ef,Watson:2012uy,Maalej:2013uu}); there still lacks a consolidated effort to capture the knowledge and artefacts best suited to \textit{manually write} API documentation.
This paper presents outcomes from a preliminary work to address this gap and offers two key contributions:
\begin{itemize}
  \item a systematic mapping study (SMS) consisting of 21 studies that capture what knowledge or artefacts should be contained within API documentation; and,
  \item a structured taxonomy based on the consolidated recommendations of these 21 studies.
\end{itemize}
After performing our SMS on what API knowledge should be captured in documentation---to assist API designers---we propose a five dimensional taxonomy consisting of: (1)~\dima{}; (2)~\dimb{}; (3)~\dimc{}; (4)~\dimd{}; and (5)~\dime{}.

This paper is structured as thus: \cref{sec:related-work} presents related work in the area; \cref{sec:method} is divided into two subsections, the first describing how primary sources were selected in a SMS, with the second describing the development of our taxonomy from these sources; \cref{sec:findings} present our primary studies and our proposed taxonomy; \cref{sec:limitations} describes the threats to validity of this work and \cref{sec:conclusions} provides concluding remarks and the future directions of this study.

\section{Related Work}
\label{sec:related-work}

Systematic mapping studies have previously been explored in the area of API usability and developer experience. \citet{Nybom:2018ef} recently performed a systematic mapping study on 36 API documentation generation tools and approaches. Presented is an analysis of state-of-the-art of the tools developed, what kind of documentation is generated by them, and the dependencies they require to generate this documentation. Their findings highlight a recent effort on the development of API documentation by producing example code snippets and/or templates on how to use the API or bootstrap developers to begin using the APIs. A secondary focus is closely followed by tools that produce natural language descriptions that can be produced within developer documentation. 
However, \citeauthor{Nybom:2018ef} produce a systematic mapping study on the types of \textit{tooling} that exists to assist in producing and validating API documentation. While this is a systematic study with key insights into the types of tooling produced, there is still a gap for a systematic mapping study in what \textit{guidelines} have been produced by the literature in developing natural-language documentation itself, which our work has addressed.

\citet{Watson:2012uy} performed a heuristic assessment of 11 high-level universal design elements of API documentation against 35 popular APIs. He demonstrated that many of these popular APIs fail to grasp even the basic of these elements; for example, 25\% of the documentation sets did not provide any basic overview documentation. However, the heuristic used within this study consists of just 11 elements and is based on only three seminal works. Our work extends these heuristics and structures them into a consolidated, hierarchical taxonomy using a systematic taxonomy development method for SE.

A taxonomy of knowledge patterns within API reference documentation by \citet{Maalej:2013uu} classified 12 distinct knowledge types. Evaluation of the taxonomy against JDK 6 and .NET 4.0 showed that, while functionality and structure of the API is well-communicated, core concepts and rationale about the API are quite rare to find. Moreover, they demonstrated that low-value `non-information' (documentation that provides uninformative boilerplate text with no insight into the API at all) is substantially present in the documentation of methods and fields in these APIs. Their findings recommend that developers factor their 12 distinct knowledge types into the process of code documentation and prevent documenting low-value documentation. The development of their taxonomy consisted of questions to model knowledge and information, thereby capturing the reason about disparate information units independent to context; a key difference to this paper is the systematic taxonomy approach utilised.

\section{Method}
\label{sec:method}

Our taxonomy development consisted of two phases. Firstly, we conducted a SMS to identify and analyse API documentation studies, following the guidelines of \citet{Kitchenham:2007dd} and \citet{Petersen:2008td}. Following this, we followed the software engineering (SE) taxonomy development method devised by \citet{Usman:2017hn} on our findings from the SMS.

\subsection{Systematic Mapping Study}
\label{sec:method:lit-review}

\subsubsection{Research Questions (RQs)}

To guide our SMS, we developed the following RQs:

\begin{enumerate}[leftmargin=4\parindent,label=\textbf{RQ\arabic*}]
  \item What knowledge do API documentation studies contribute?
  \item How is API documentation studied?
\end{enumerate}

The intent behind RQ1 is to collate as much of the insight provided by the literature on how API providers should best document their work. This helped us shape and form the taxonomy provided in \cref{sec:findings}. RQ2 addresses methodologies by which these studies come to these conclusions to identify gaps in literature where future studies can potentially focus.


\subsubsection{Automatic Filtering}

Informed by similar previous studies in SE  \citep{Glass:2002wa,Usman:2017hn,GAROUSI2019101}, we begin by defining the SWEBOK \citep{IEEE:1990wp} knowledge areas (KAs) to assist in the search and mapping process of an SMS. Our search query was built using related KAs, relevant synonyms, and the term `software engineering' (for comprehensiveness) all joined with the OR operator. Due to the lack of a standard definition of an API, we include the terms: `API' and its expanded term; software library, component and framework; and lastly SDK and its expanded term. These too were joined with the OR operator, appended with an AND. Lastly, the term `documentation' was appended with an AND.
Our final search string was:
\begin{framed}
\noindent
\parbox{\linewidth}{
\footnotesize
( ``software design'' \textbf{OR} ``software architecture" \textbf{OR} ``software construction" \textbf{OR} ``software development" \textbf{OR} ``software maintenance" \textbf{OR} ``software engineering process" \textbf{OR} ``software process" \textbf{OR} ``software lifecycle" \textbf{OR} ``software methods" \textbf{OR} ``software quality" \textbf{OR} ``software engineering professional practice" \textbf{OR} ``software engineering" ) \textbf{AND} ( api \textbf{OR} ``application programming interface" \textbf{OR} ``software library" \textbf{OR} ``software component" \textbf{OR} ``software framework" \textbf{OR} sdk \textbf{OR} ``software development kit" ) \textbf{AND} ( documentation )
}
\end{framed}

The query was then executed on all available metadata (title, abstract and keywords) on three primary sources to search for relevant studies in May 2019. Web of Science\footnoteurl{http://apps.webofknowlegde.com}{23 May 2019}  (WoS), Compendex/Inspec\footnoteurl{http://www.engineeringvillage.com}{23 May 2019} (C/I) and Scopus\footnoteurl{http://www.scopus.com}{23 May 2019} were chosen due to their relevance in SE literature (containing the IEEE, ACM, Springer and Elsevier databases) and their ability to support advanced queries \citep{Brereton:2007by,Kitchenham:2007dd}. A total 4,501 results\footnote{Raw results can be located at \url{http://bit.ly/2KxBLs4}} were found, with 549 being duplicates. \Cref{tab:search-results} displays our results in further detail (duplicates not omitted).

\begin{table}[tb]
  \caption{Summary of our search results and publication types}
  \label{tab:search-results}
  \centering
  \begin{tabular}{|l||lll||l|}
    \hline
    \textbf{Publication type} &
    \textbf{WoS} &
    \textbf{C/I} &
    \textbf{Scopus} &
    \textbf{Total} \\
    \hline
    \hline
    Conference Paper & 27 & 442 & 2353 & 2822 \\
    Journal Article & 41 & 127 & 1236 & 1404\\
    Book & 23 & 17 & 224 & 264\\
    Other & 0 & 5 & 6 & 11\\
    \hline
    \textbf{Total} & 91 & 591 & 3819 & 4501\\
    \hline
  \end{tabular}
\end{table}



\subsubsection{Manual Filtering}

A follow-up manual filtering to select primary studies was performed on the 4,501 results using the following inclusion criteria (IC) and exclusion criteria (EC):

\begin{enumerate}[leftmargin=4\parindent,label=\textbf{IC\arabic*}]
  \item Studies must be relevant to API documentation: specifically, we exclude studies that deal with improving the technical API usability (e.g., improved usage patterns);
  \item Studies must propose new knowledge or recommendations to document APIs;
  \item Studies must be relevant to SE as defined in SWEBOK;
\end{enumerate}
\begin{enumerate}[leftmargin=4\parindent,label=\textbf{EC\arabic*}]
  \item Studies where full-text is not accessible through standard institutional databases; 
  \item Studies that do not propose or extend how to improve the official, natural language documentation of an API;
  \item Studies proposing a third-party tool to enhance existing documentation or generate new documentation using data mining (i.e., not proposing strategies to improve official documentation);
  \item Studies not written in English;
  \item Studies not peer-reviewed.
\end{enumerate}
\smallskip

After exporting metadata of search results to a spreadsheet, a three-phase curation process was conducted. The first author read the publication source (to omit non-SE papers quickly), author keywords and title of all 4,501 studies (514 that were duplicates), and abstract. As we considered multiple databases, some studies were repeated. However, the DOIs and titles were sorted and reviewed, retaining only one copy of the paper from a single database. Moreover, as there was no limit to the year range in our query, some studies were republished in various venues. These, too, were handled with title similarity matching, wherein only the first paper was considered. Where the inclusion or exclusion criteria could not be determined from the abstract alone, the paper was automatically shortlisted. Any doubt in a study automatically included it into the second phase. This resulted in 133 studies being shortlisted to the second phase. We rejected 427 studies that were unrelated to SE, 3,235 were not directly related to documenting APIs (e.g., to enhance coding techniques that improve the overall developer usability of the API), 182 proposed new tools to enhance API documentation or used machine learning to mine developer's discussion of APIs, and 10 were not in English.

The shortlisted studies were then re-evaluated by re-reading the abstract, the introduction and conclusion. Performing this second phase removed a further 64 studies that were on API usability or non API-related  documentation (i.e., code commenting); we further refined our exclusion criteria to better match the research outcomes of this goal (chiefly including the word `natural language' documentation in EC2) which removed studies focused to improve technical documentation of APIs such as data types and communication schemas. Additionally, 26 studies were removed as they were related to introducing new tools (EC3), 3 were focused on tools to mine API documentation, 7 studies where no recommendations were provided, 2 further duplicate studies, and a further 10 studies where the full text was not available, not peer reviewed or in English. Books are commonly not peer-reviewed (EC5), however no books were shortlisted within these results. \textbf{This resulted in 21 primary studies for further analysis. The mapping of primary study identifiers to references S1--21 can be found at \url{http://bit.ly/2MtsIuE}}.

Intra-rater reliability of our 133 shortlisted papers was tested using the test-retest approach \citep{Kitchenham:2007dd} by re-evaluating a random sample of 10\% (13 total) of the studies shortlisted above a week after initial studies were shortlisted. Using the Cohen's kappa coefficient as a metric for reliability, $\kappa=0.7547$, indicating substantial agreement \citep{Landis:1977kv}.

\subsubsection{Data Extraction}
\label{sec:data-extraction}

Of the 21 primary studies, we conducted abstract key-wording adhering to \citeauthor{Petersen:2008td}'s guidelines \citep{Petersen:2008td} to develop a classification scheme.
An initial set of keywords were applied for each paper in terms of their methodologies and research approaches (RQ2), based on an existing classification schema by \citet{Wieringa:2006vd}: evaluation, validation, personal experience and philosophical papers.

After all primary studies had been assigned keywords, we noticed that \textbf{all papers used field study techniques}, and thus we consolidated these keywords using \citeauthor{Singer:2007tu}'s framework of SE field study techniques \citep{Singer:2007tu}. \citeauthor{Singer:2007tu} captures both study techniques \textit{and} methods to collect data within the one framework, namely: \textit{direct techniques}, including brainstorming and focus groups, interviews and questionnaires, conceptual modelling, work diaries, think-aloud sessions, shadowing and observation, participant observation; \textit{indirect techniques}, including instrumenting systems, fly-on-the-wall; and \textit{independent techniques}, including analysis of work databases, tool use logs, documentation analysis, and static and dynamic analysis. 

\begin{figure}[bt]
  \centering
  \includegraphics[width=\linewidth]{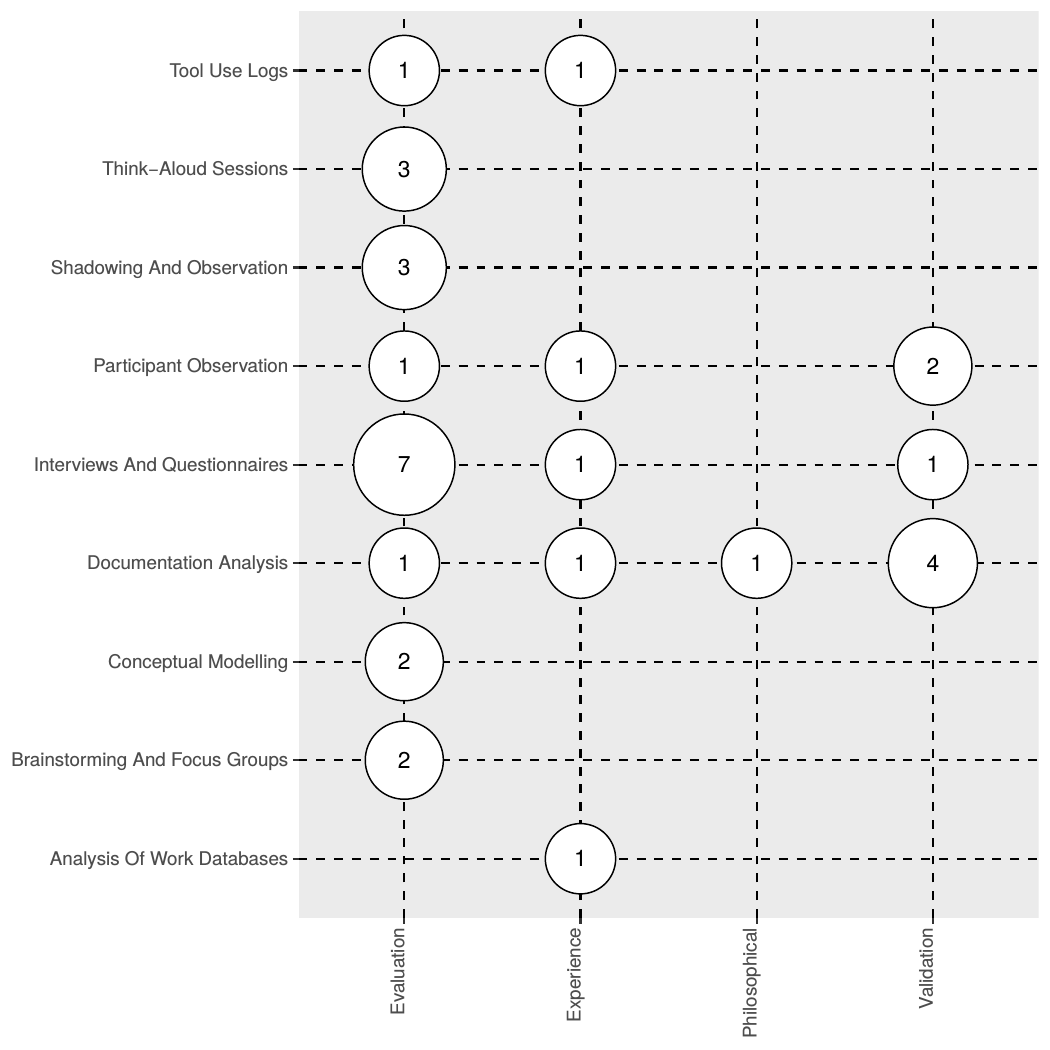}
  \caption{Systematic map - field study technique vs research type}
  \label{fig:sms}
\end{figure}

\Cref{tab:extraction} describes our data extraction form, which was used to collect relevant data from each paper. \Cref{fig:sms} maps each study to one (or more, if applicable) of methodologies plotted against \citeauthor{Wieringa:2006vd}'s research approaches.

\begin{table}[tb]
  \caption{Data extraction form}
  \label{tab:extraction}
  \centering
  \begin{tabular}{|l|p{0.6\linewidth}|}
    \hline
    \textbf{Data item(s)} &
    \textbf{Description}
    \\
    \hline
    \hline
    Citation metadata & Title, author(s), years, publication venue, publication type \\
    Key recommendation(s) & As per IC2, the study must propose at least one recommendation on what should be captured in API documentation \\
    Evaluation method & Did the authors evaluate their recommendations? If so, how? \\
    Primary technique & The primary technique used to devise the recommendation(s) \\ 
    Secondary technique & As above, if a second study was conducted \\
    Tertiary technique & As above, if a third study was conducted \\
    Research type & The research type employed in the study as defined by \citeauthor{Wieringa:2006vd}'s taxonomy \\
    \hline
  \end{tabular}
\end{table}

\subsection{Development of the Taxonomy}
\label{sec:method:taxonomy-development}

\citeauthor{Usman:2017hn} concludes that a majority of SE taxonomies are developed in an ad-hoc way \citep{Usman:2017hn}, and proposes a systematic approach to develop taxonomies in SE that extends previous efforts by including lessons learned from more mature fields. In this subsection, we outline the 4 phases and 13 steps taken to develop our taxonomy based on \citeauthor{Usman:2017hn}'s technique.

\subsubsection{Planning phase} The planning phase of the technique involves the following six steps:

\begin{enumerate}[label=\textbf{(\arabic*)}]
  \item \textit{defining the SE KA}: The software engineering KA, as defined by the SWEBOK, is software construction;
  \item \textit{defining the objective}: The main objective of the proposed taxonomy is to define a set of categories that enables to classify different facets of natural-language API \textit{documentation} knowledge (not API \textit{usability} knowledge) as reported in existing literature;
  \item \textit{defining the subject matter}: The subject matter of our proposed taxonomy is  documentation artefacts of APIs;
  \item \textit{defining the classification structure}: The classification structure of our  proposed taxonomy is \textit{hierarchical};
  \item \textit{defining the classification procedure}: The procedure used to classify the documentation artefacts is qualitative; 
  \item \textit{defining the data sources}: The basis of the taxonomy is derived from field study techniques (see \cref{sec:data-extraction}).
\end{enumerate}

\subsubsection{Identification and extraction phase} The second phase of the taxonomy development involves \textbf{(7)}~\textit{extracting all terms and concepts} from relevant literature, as selected from our 21 primary studies. These terms are then consolidated by \textbf{(8)}~\textit{performing terminology control}, as some terms may refer to different concepts and vice-versa.

\subsubsection{Design phase} The design phase identified the core dimensions and categories within the extracted data items. The first step is to \textbf{(9)}~\textit{identify and define taxonomy dimensions}; for this study we utilised a bottom-up approach to identify the dimensions, i.e. extracting the categories first and then nominating which dimensions these categories fit into using an iterative approach. As a bottom-up approach was utilised, step (9) also encompassed the second stage of the design phase, which is to \textbf{(10)}~\textit{identify and describe the categories} of each dimension. Thirdly, we \textbf{(11)}~\textit{identify and describe relationships} between dimensions and categories, which can be skipped if the relationships are too close together, as is the case of our grouping technique which allows for new dimensions and categories to be added. The last step in this phase is to \textbf{(12)}~\textit{define guidelines for using and updating the taxonomy}, however as this taxonomy still an emerging result, guidelines to update and use the taxonomy are anticipated future work.

\subsubsection{Validation phase} In the final phase of taxonomy development, taxonomy designers must \textbf{(13)}~\textit{validate the taxonomy} to assess its usefulness. Ideally, this is done by applying the taxonomy heuristically against developers or real-world case-studies. This remains a plan for future work (see \cref{sec:limitations}).

\def\cn{}
\def\cy{\checkmark}

\begin{table*}[hbt]
  \caption{An overview of the 5 dimensions and categories (sub-dimensions) within our proposed taxonomy.}
  \label{tab:taxonomy}
  \begin{tabular}{|rp{0.58\linewidth}||p{0.2\linewidth}|c|}
    \hline

    \textbf{Key} &
    \textbf{Description: Dimensions A=\dima{}; B=\dimb{}; C=\dimc{}; D=\dimd{}; E=\dime{}} &
    \textbf{Primary Studies } &
    \textbf{Total (\%)} \\

    \hline
    \hline
    [A1]&
    Quick-start guides to rapidly get started using the API in a specific programming language.
    &
    [S4, S9, S10] &
    3/21 (14\%)\\

    \hline
    \textbf{[A2]}&
    \textbf{Low-level reference manual documenting all API components to review fine-grade detail.}
    &
    \textbf{[S1, S3, S4, S8, S9, S10, S11, S12, S15, S16, S17]} &
    \textbf{11/21 (52\%)}\\

    \hline
    [A3]&
    Explanations of the API's high-level architecture to better understand intent and context.
    &
    [S1, S2, S4, S11, S14, S16, S19, S20] &
    8/21 (38\%)\\

    \hline
    [A4]&
    Source code implementation and code comments (where applicable) to understand the API author's mindset.
    &
    [S1, S4, S7, S12, S13, S17, S20] &
    7/21 (33\%)\\

    \hline
    \textbf{[A5]}&
    \textbf{Code snippets (with comments) of no more than 30 LoC to understand a basic component functionality within the API.}
    &
    \textbf{[S1, S2, S4, S5, S6, S7, S9, S10, S11, S14, S15, S16, S18, S20, S21]} &
    \textbf{15/21 (71\%)}\\

    \hline
    \textbf{[A6]}&
    \textbf{Step-by-step tutorials, with screenshots to understand  how to build a non-trivial piece of functionality with multiple components of the API.}
    &
    \textbf{[S1, S2, S4, S5, S7, S9, S10, S15, S16, S18, S20, S21]} &
    \textbf{12/21 (57\%)}\\

    \hline
    [A7]&
    Downloadable source code of production-ready applications that use the API to understand implementation in a large-scale solution.
    &
    [S1, S2, S5, S9, S15] &
    5/21 (24\%)\\

    \hline
    [A8]&
    Best-practices of implementation to assist with debugging and efficient use of the API.
    &
    [S1, S2, S4, S5, S7, S8, S9, S14] &
    8/21 (38\%)\\

    \hline
    [A9]&
    An exhaustive list of all major components that exist within the API.
    &
    [S4, S16, S19] &
    3/21 (14\%)\\

    \hline
    [A10]&
    Minimum system requirements and dependencies to use the API.
    &
    [S4, S7, S13, S17, S19] &
    5/21 (24\%)\\

    \hline
    [A11]&
    Instructions to install or begin using the API and details on its release cycle and updating it.
    &
    [S4, S7, S8, S9, S11, S13, S16, S19] &
    8/21 (38\%)\\

    \hline
    [A12]&
    Error definitions that describe how to address a specific problem.
    &
    [S1, S2, S4, S5, S9, S11, S13] &
    7/21 (33\%)\\

    \hline
    \hline
    \textbf{[B1]}&
    \textbf{A brief description of the purpose or overview of the API as a low barrier to entry.}
    &
    \textbf{[S1, S2, S4, S5, S6, S8, S10, S11, S15, S16]} &
    \textbf{10/21 (48\%)}\\

    \hline
    [B2]&
    Descriptions of the types of applications the API can develop.
    &
    [S2, S4, S9, S11, S15, S18] &
    6/21 (29\%)\\

    \hline
    [B3]&
    Descriptions of the types of users who should use the API.
    &
    [S4, S9] &
    2/21 (10\%)\\

    \hline
    [B4]&
    Descriptions of the types of users who will use the product the API creates.
    &
    [S4] &
    1/21 (5\%)\\

    \hline
    [B5]&
    Success stories about the API used in production.
    &
    [S4] &
    1/21 (5\%)\\

    \hline
    [B6]&
    Documentation to compare similar APIs within the context to this API.
    &
    [S2, S6, S13, S18] &
    4/21 (19\%)\\

    \hline
    [B7]&
    Limitations on what the API can and cannot provide.
    &
    [S4, S5, S8, S9, S14, S16] &
    6/21 (29\%)\\

    \hline
    \hline
    [C1]&
    Descriptions of the relationship between API components and domain concepts.
    &
    [S3, S10] &
    2/21 (10\%)\\

    \hline
    [C2]&
    Definitions of domain-terminology and concepts, with synonyms if applicable.
    &
    [S2, S3, S4, S6, S7, S10, S14, S16] &
    8/21 (38\%)\\

    \hline
    [C3]&
    Generalised documentation for non-technical audiences regarding the API and its domain.
    &
    [S4, S8, S16] &
    3/21 (14\%)\\

    \hline
    \hline
    [D1]&
    A list of FAQs.
    &
    [S4, S7] &
    2/21 (10\%)\\

    \hline
    [D2]&
    Troubleshooting suggestions.
    &
    [S4, S8] &
    2/21 (10\%)\\

    \hline
    [D3]&
    Diagrammatically representing API components using visual architectural representations.
    &
    [S6, S13, S20] &
    3/21 (14\%)\\

    \hline
    [D4]&
    Contact information for technical support.
    &
    [S4, S8, S19] &
    3/21 (14\%)\\

    \hline
    [D5]&
    A printed/printable resource for assistance.
    &
    [S4, S6, S7, S9, S16] &
    5/21 (24\%)\\

    \hline
    [D6]&
    Licensing information.
    &
    [S7] &
    1/21 (5\%)\\

    \hline
    \hline
    [E1]&
    Searchable knowledge base.
    &
    [S3, S4, S6, S10, S14, S17, S18] &
    7/21 (33\%)\\

    \hline
    [E2]&
    Context-specific discussion forum.
    &
    [S4, S10, S11] &
    3/21 (14\%)\\

    \hline
    [E3]&
    Quick-links to other relevant documentation frequently viewed by developers.
    &
    [S6, S16, S20] &
    3/21 (14\%)\\

    \hline
    [E4]&
    Structured navigational style (e.g., breadcrumbs).
    &
    [S6, S10, S20] &
    3/21 (14\%)\\

    \hline
    [E5]&
    Visualised map of navigational paths to certain API components in the website.
    &
    [S6, S14, S20] &
    3/21 (14\%)\\

    \hline
    \textbf{[E6]}&
    \textbf{Consistent look and feel of documentation.}
    &
    \textbf{[S1, S2, S3, S5, S6, S8, S10, S15, S20]} &
    \textbf{9/21 (43\%)}\\
    \hline
  \end{tabular}
\end{table*}

\section{Taxonomy}
\label{sec:findings}

Our taxonomy consists of five dimensions (labelled A--E) that respectively cover:
\textbf{[A] \dima{}} on \textit{how} to use the API for the developer's intended use case;
\textbf{[B] \dimb{}} on \textit{when} the developer should choose this API for a particular use case;
\textbf{[C] \dimc{}} of the domain behind the API to understand \textit{why} this API should be chosen for this domain;
\textbf{[D] \dimd{}} that describe \textit{what} additional documentation the API provides; and
 \textbf{[E] \dime{}} to help organise the \textit{visualisation} of the above information.
Further descriptions of the categories encompassing each dimension are given within \cref{tab:taxonomy}, coded as [$Xi$], where $i$ is the category identifier within a dimension, $X$, where $X~\in~\{ A, B, C, D, E \}$.

\textbf{We expand these five dimensions into 34 categories (sub-dimensions) and \cref{tab:taxonomy} provides a weighting of these categories in the rightmost column as calculated as a percentage of the number of primary studies per category divided by the total of primary studies.} The top five weighted categories (bolded in \cref{tab:taxonomy}) highlight what most studies recommend documenting in API documentation, with the top three falling under the \dima{} dimension.

The majority (71\%) of studies advocate for \textbf{code snippets} as a necessary piece in the API documentation puzzle [A5]. While code snippets generally only reflect small portions of API functionality (limited to 15--30 LoC), this is complimented by \textbf{step-by-step tutorials} (57\% of studies) that tie in multiple (disparate) components of API functionality, generally with some form of screenshots, demonstrating the development of a non-trivial application using the API step-by-step [A6]. The third highest category weighted was also under the \dima{} dimension, being \textbf{low-level reference documentation} at 52\% [A2]. These three categories were the only categories to be weighted as majority categories (i.e., their weighting was above 50\%).

The fourth and fifth highest weights are \textbf{an entry-level purpose/overview of the API} (48\%) that gives a brief motivation as to why a developer should choose a particular API over another [B1] and \textbf{consistency in the look and feel} of the documentation throughout all of the API's official documentation (43\%) [E6].



 
\section{Threats to Validity}
\label{sec:limitations}

Threats to \textit{internal validity} concern factors internal to our study that may affect results. Guidelines on producing systematic reviews \citep{Kitchenham:2007dd} suggest that single researchers conducting their reviews should discuss the review protocol, inclusion decisions, data extraction with a third party. In this paper, we have presented the early outcomes of our systematic review, which has utilised the test-retest methodology as a measure of reliability. \citet{5416726} states that a defining characteristic of any SMS is to test the reliability of the review and extraction processes. We plan to mitigate this threat by conducting \textit{inter}-relater reliability with the continuation of this work, using independent analysis and conflict resolution as per guidelines suggested by \citet{Garousi:2017:EGE:3084226.3084238}. Similarly, the development of our taxonomy would benefit from an inter-rater reliability categorisation of a sample of papers to both ensure that our weightings of categories are reliable and that the categories and dimensions fit the objectives of the taxonomy. Furthermore, a future user study (see \cref{sec:conclusions}) will be needed to assess whether the extracted information from API documentation actually impacts on developer productivity, and the usefulness of such a taxonomy should be evaluated.


Threats to \textit{external validity} represent the generalisation of the observations we have found in this study. While we have used a broad range of literature that encompasses API documentation guidelines, we acknowledge that not all papers contributing to API documentation may have been captured in the taxonomy. All efforts were made to include as many papers as possible given our filtering technique, though it is likely that some papers filtered out (e.g., papers not in English) may alter our conclusions, introducing conflicting recommendations. However, given the consistency of these trends within the studies that were sourced, we consider this a low likelihood.

Threats to \textit{construct validity} relates to the degree by which the data extrapolated in this study sufficiently measures its intended goals. Automatic searching was conducted in the SMS by choice of three popular databases (see \cref{sec:method:lit-review}). As a consequence of selecting multiple databases, duplicates were returned. This was mitigated by manually curating out all duplicate results from the set of studies returned. Additionally, we acknowledge that the lack manual searching of papers within particular venues may be an additional threat due to the misalignment of search query keywords to intended papers of inclusion. Thus, our conclusions are only applicable to the information we were able to extract and summarise, given the primary sources selected.

\section{Conclusions \& Future Work}
\label{sec:conclusions}

API documentation is an aspect of quality of software, as it facilitates the developer's productivity and assists with evolution. Improving the quality of the documentation of third party APIs improves the quality of software using them.

To date, we did not find a systematic literature review that offers a consolidated taxonomy of key recommendations. Moreover, there has been little work on mapping the research produced in this space against the techniques used to arrive at the recommendations.
Starting with 4,501 papers potentially relating to API documentation, we identified 21 key relevant studies, and synthesise a taxonomy of the various documentation aspects that should improve API documentation quality. Furthermore, we also capture the most commonly used analysis techniques used in the academic literature. \Cref{fig:sms} highlights that a majority of these studies employ interviews and questionnaires, and only some undertake structured documentation analysis. 

In future revisions of this work, we intend use our results as the input to a restricted systematic literature review in API documentation artefacts. In doing so, we will consider conducting the following:

\begin{itemize}
  \item improving reliability metrics of our study (see \cref{sec:limitations}) with an inter-rater reliability method;
  \item the development and applicability of our taxonomy will be further explored by triangulating the taxonomy  against actual developers in industry to assess the efficacy of these recommendations---this will empirically reflect what is important from a \textit{practitioner} point of view;
  \item reviewing the techniques and evaluation of our selected studies to extract the effectiveness of the various approaches used in the conclusions;
  \item conducting a heuristic validation of the taxonomy against intelligent APIs, given the current trend in SE that is exploring how machine learning and artificial intelligence-based applications may affect existing approaches;
  \item  arrive at a relevance ranking for each of the 34 categories, based on developer surveys and current weights.
\end{itemize}

We believe the results of this preliminary empirical work may provide further insight for future follow-up user studies with developers. Whilst our aim is to eventually improve the quality of API documentation, the ultimate goal is  improving the developer's experience when producing systems and, therefore, improving the efficacy and productivity at which software is produced within industry.   We hope that API designers will utilise the taxonomy produced in this paper as a weighted checklist for what should be considered in their own APIs.


\balance
\bibliography{papers}

\end{document}